\documentclass[letterpaper,twocolumn,prl,aps,superscriptaddress,floatfix]{revtex4}
\usepackage[latin1]{inputenc}
\usepackage{bm}
\usepackage{amssymb,amsbsy,amsmath}
\usepackage{multirow,amssymb,amsbsy,amsmath}
\usepackage{graphicx}
\usepackage{verbatim}
\usepackage{color}
\usepackage{pifont}
\usepackage{hyperref}
\usepackage{xcolor}
\usepackage{float}

\begin{document}

\title{A universal programmable Gaussian Boson Sampler for drug discovery}
\author{Shang Yu$\footnote{These authors contributed equally to this work}$}
\email{shang.yu@imperial.ac.uk}
\affiliation{Research Center for Quantum Sensing, Zhejiang Lab, Hangzhou, 310000, People's Republic of China}
\affiliation{Quantum Optics and Laser Science, Blackett Laboratory, Imperial College London, Prince Consort Rd, London SW7 2AZ, United Kingdom}
\affiliation{CAS Key Laboratory of Quantum Information, University of Science and Technology of China, Hefei, Anhui 230026, China}
\affiliation{CAS Center For Excellence in Quantum Information and Quantum Physics,
University of Science and Technology of China, Hefei, 230026, China}

\author{Zhi-Peng Zhong$^*$}
\affiliation{Research Center for Quantum Sensing, Zhejiang Lab, Hangzhou, 310000, People's Republic of China}

\author{Yuhua Fang$^*$}
\affiliation{College of Pharmacy, Faculty of Health Science, University of Manitoba, Winnipeg, MB R3E 0T6, Canada}

\author{Raj B. Patel}
\email{raj.patel1@imperial.ac.uk}
\affiliation{Quantum Optics and Laser Science, Blackett Laboratory, Imperial College London, Prince Consort Rd, London SW7 2AZ, United Kingdom}

\author{Qing-Peng Li}
\affiliation{Research Center for Quantum Sensing, Zhejiang Lab, Hangzhou, 310000, People's Republic of China}

\author{Wei Liu}
\affiliation{CAS Key Laboratory of Quantum Information, University of Science and Technology of China, Hefei, Anhui 230026, China}
\affiliation{CAS Center For Excellence in Quantum Information and Quantum Physics,
University of Science and Technology of China, Hefei, 230026, China}

\author{Zhenghao Li}
\affiliation{Quantum Optics and Laser Science, Blackett Laboratory, Imperial College London, Prince Consort Rd, London SW7 2AZ, United Kingdom}

\author{Liang Xu}
\affiliation{Research Center for Quantum Sensing, Zhejiang Lab, Hangzhou, 310000, People's Republic of China}

\author{Steven Sagona-Stophel}
\affiliation{Quantum Optics and Laser Science, Blackett Laboratory, Imperial College London, Prince Consort Rd, London SW7 2AZ, United Kingdom}

\author{Ewan Mer}
\affiliation{Quantum Optics and Laser Science, Blackett Laboratory, Imperial College London, Prince Consort Rd, London SW7 2AZ, United Kingdom}

\author{Sarah E. Thomas}
\affiliation{Quantum Optics and Laser Science, Blackett Laboratory, Imperial College London, Prince Consort Rd, London SW7 2AZ, United Kingdom}

\author{Yu Meng}
\affiliation{CAS Key Laboratory of Quantum Information, University of Science and Technology of China, Hefei, Anhui 230026, China}
\affiliation{CAS Center For Excellence in Quantum Information and Quantum Physics,
University of Science and Technology of China, Hefei, 230026, China}

\author{Zhi-Peng Li}
\affiliation{CAS Key Laboratory of Quantum Information, University of Science and Technology of China, Hefei, Anhui 230026, China}
\affiliation{CAS Center For Excellence in Quantum Information and Quantum Physics,
University of Science and Technology of China, Hefei, 230026, China}

\author{Yuan-Ze Yang}
\affiliation{CAS Key Laboratory of Quantum Information, University of Science and Technology of China, Hefei, Anhui 230026, China}
\affiliation{CAS Center For Excellence in Quantum Information and Quantum Physics,
University of Science and Technology of China, Hefei, 230026, China}

\author{Zhao-An Wang}
\affiliation{CAS Key Laboratory of Quantum Information, University of Science and Technology of China, Hefei, Anhui 230026, China}
\affiliation{CAS Center For Excellence in Quantum Information and Quantum Physics,
University of Science and Technology of China, Hefei, 230026, China}

\author{Nai-Jie Guo}
\affiliation{CAS Key Laboratory of Quantum Information, University of Science and Technology of China, Hefei, Anhui 230026, China}
\affiliation{CAS Center For Excellence in Quantum Information and Quantum Physics,
University of Science and Technology of China, Hefei, 230026, China}

\author{Wen-Hao Zhang}
\affiliation{CAS Key Laboratory of Quantum Information, University of Science and Technology of China, Hefei, Anhui 230026, China}
\affiliation{CAS Center For Excellence in Quantum Information and Quantum Physics,
University of Science and Technology of China, Hefei, 230026, China}

\author{Geoffrey K Tranmer}
\affiliation{College of Pharmacy, Faculty of Health Science, University of Manitoba, Winnipeg, MB R3E 0T6, Canada}

\author{Ying Dong}
\affiliation{Research Center for Quantum Sensing, Zhejiang Lab, Hangzhou, 310000, People's Republic of China}

\author{Yi-Tao Wang}
\email{yitao@ustc.edu.cn}
\affiliation{CAS Key Laboratory of Quantum Information, University of Science and Technology of China, Hefei, Anhui 230026, China}
\affiliation{CAS Center For Excellence in Quantum Information and Quantum Physics,
University of Science and Technology of China, Hefei, 230026, China}

\author{Jian-Shun Tang}
\email{tjs@ustc.edu.cn}
\affiliation{CAS Key Laboratory of Quantum Information, University of Science and Technology of China, Hefei, Anhui 230026, China}
\affiliation{CAS Center For Excellence in Quantum Information and Quantum Physics,
University of Science and Technology of China, Hefei, 230026, China}
\affiliation{Hefei National Laboratory, University of Science and Technology of China, Hefei 230088, China}

\author{Chuan-Feng Li}
\email{cfli@ustc.edu.cn}
\affiliation{CAS Key Laboratory of Quantum Information, University of Science and Technology of China, Hefei, Anhui 230026, China}
\affiliation{CAS Center For Excellence in Quantum Information and Quantum Physics,
University of Science and Technology of China, Hefei, 230026, China}
\affiliation{Hefei National Laboratory, University of Science and Technology of China, Hefei 230088, China}

\author{Ian A. Walmsley}
\affiliation{Quantum Optics and Laser Science, Blackett Laboratory, Imperial College London, Prince Consort Rd, London SW7 2AZ, United Kingdom}

\author{Guang-Can Guo}
\affiliation{CAS Key Laboratory of Quantum Information, University of Science and Technology of China, Hefei, Anhui 230026, China}
\affiliation{CAS Center For Excellence in Quantum Information and Quantum Physics,
University of Science and Technology of China, Hefei, 230026, China}
\affiliation{Hefei National Laboratory, University of Science and Technology of China, Hefei 230088, China}

\begin{abstract}
{Gaussian Boson Sampling (GBS) has the potential to solve complex graph problems, such as clique-finding, which is relevant to drug discovery tasks. However, realizing the full benefits of quantum enhancements requires a large-scale quantum hardware with universal programmability. Here, we have developed a time-bin encoded GBS photonic quantum processor that is universal, programmable, and software-scalable. Our processor features freely adjustable squeezing parameters and can implement arbitrary unitary operations with a programmable interferometer. Leveraging our processor, we successfully executed clique-finding on a 32-node graph, achieving approximately twice the success probability compared to classical sampling. Additionally, we established a versatile quantum drug discovery platform using this GBS processor, enabling molecular docking and RNA folding prediction tasks. Our work achieves the state-of-the-art in GBS circuitry with its distinctive universal and programmable architecture which advances GBS towards real-world applications.}
\end{abstract}

\maketitle
\date{\today}
Quantum computing technology has developed rapidly in recent years~\cite{Spring2013,Barends2016,Mohseni2017,Arute2019,Zhong2020,Arrazola2021,Madsen2022}, and an exponential ``speed-up'' compared to classical methods has been experimentally demonstrated for certain algorithms~\cite{Zhang2017,Bernien2017,Arute2019,Zhong2020,Madsen2022}. Quantum sampling tasks, like boson sampling~\cite{Tillmann2013,Brod2019,Lund2014}, have proven to be challenging to solve on classical computers within a reasonable time frame, but can be implemented and solved efficiently on photonic processors~\cite{Spring2013,Wang2017}. As a variant of boson sampling, Gaussian Boson sampling (GBS)~\cite{Hamilton2017} uses squeezed light as the input states making it easier to scale and therefore shows great capacity to demonstrate quantum advantage in optical systems~\cite{Zhong2020,Madsen2022}.

The prospect of achieving quantum advantage has motivated the discovery of several real-world applications, such as dense graph searching~\cite{Arrazola2018,Sempere2022}, molecular vibronic spectra calculations~\cite{Huh2015,Arrazola2021}, and molecular docking~\cite{Banchi2020}.
In these tasks, a GBS device should be programmable and scalable to a large number of modes~\cite{Arrazola2021,Zhong2020}. However, it is a challenging task~\cite{Sempere2022} due to the experimental complexity involved in preparing a large number of individually addressable input states and phase-shifters to achieve universal programmability~\cite{Zhong2020,Arrazola2021}.

\begin{figure*}[hbt]
\centering
\includegraphics[width=0.99\textwidth]{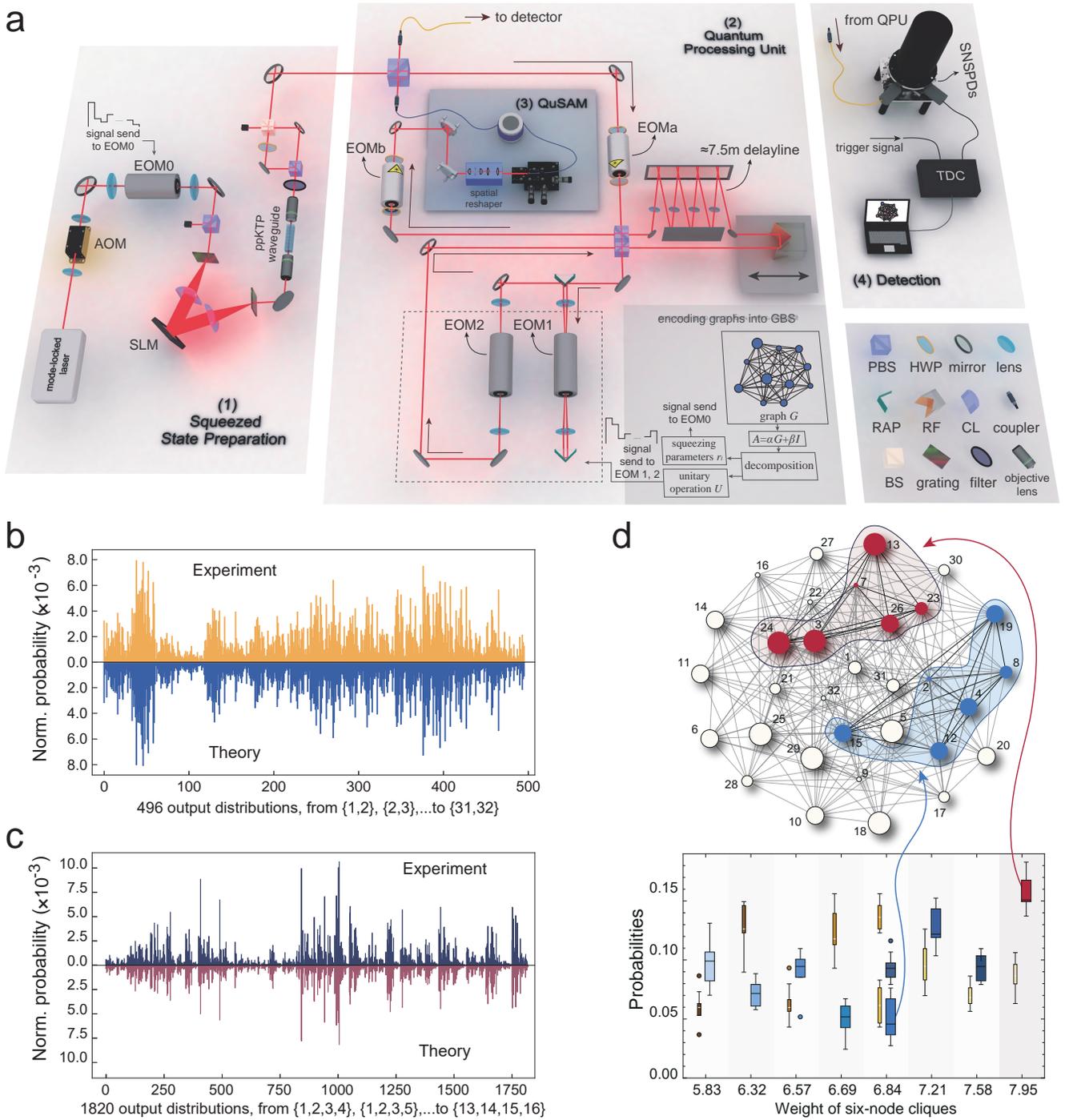}
\caption{\textbf{a,} \textbf{Universal programmable time-bin encoded GBS machine}. The GBS machine consists of four main parts: (1) Squeezed state preparation, (2) Quantum processing unit, (3) Quantum sequential access memory (QuSAM), and (4) Detection. The control system including three arbitrary wave generators is omitted for clarity. Abbreviations: PBS: polarization beam splitter, HWP: half-wave plate, RAP: right-angle prism mirrors, RF: roof prism mirror, CL: cylindrical lens, BS: beam splitter.
See Methods and Supplementary Information for details.
\textbf{b} and \textbf{c,} \textbf{Gaussian boson sampling results} \textbf{b,} Probability distribution of all 496 two-photon detection events in a 32-mode experiment. \textbf{c,} Probability distribution of all 1820 four-photon detection events in a 16-mode experiment. The horizontal axis labels output distributions in increasing order $\{(1, 2), (1, 3), \dots, (31, 32)\}$ or $\{(1, 2, 3, 4), (1, 2, 3, 5), \dots, (13, 14, 15, 16)\}$ from left to right
\textbf{d,} \textbf{Finding the maximum weighted clique in a 32-node graph}. The normalized average probabilities of the six-node cliques in the graph $\mathcal{G}_{32}$ is shown at the bottom with the corresponding graph shown above. Labels beside the nodes denote the corresponding order, and the weight of the nodes is represented by their size. The probabilities are calculated from ten individual experiments each with around 300 samples. The bars represent the GBS experimental results, and the squares are the corresponding classical uniform sampling data. Evidently, the maximum weighted clique can be found with higher success probability by using GBS machine (shown as the red bar). The error-bars are obtained from standard deviations.
}
\label{Fig1}
\end{figure*}

Time-bin encoding of Gaussian states is an effective means of achieving scale and programmability~\cite{Motes2014,He2017,Deshpande2022,Sempere2022,Madsen2022}. First, it is resource efficient where only one squeezed source and one detector are required~\cite{Sempere2022}. Second, time-bin operation provides phase stability and exhibits comparable losses with other approaches~\cite{Qi2018}. Furthermore, time-bin interferometers shows flexibility in reconfiguration since it can realize arbitrary-dimension linear transformations with the same setup. Recently, quantum computational advantage with a programmable time-bin-encoded GBS~\cite{Madsen2022} machine has been demonstrated albeit whilst sacrificing universality to avoid the accumulation of loss.

This prompts us to consider a universal and programmable time-bin GBS machine that can fulfill various practical tasks. Besides, the GBS algorithm can potentially be applied to many important problems and enhance their performance, for example, the complete subgraph (clique) finding task~\cite{Plummer1993,Bradler2018}. Some structural-based drug design methods, like molecular docking or protein folding prediction, can be interpreted as such a problem of finding the maximum weighted clique in their corresponding graph models~\cite{Banchi2020,Karp1972,Tang2022}. This indicates that a universal programmable GBS machine equipped with freely adjustable squeezers and interferometer can be utilized for the above tasks and extend the range of practical applications based on graph theory. Inspired by this prospect, we built a scalable, universal, and programmable time-bin GBS machine in this work, and make a significant stride towards using GBS in drug discovery applications.

\emph{Programmable GBS machine and sampling results}---The GBS machine shown in Fig. 1, called \emph{Abacus}, can be divided into four main parts which we now describe.
(1) Tunable squeezed state source. The pump light from a mode-locked pulsed laser (80MHz, 773nm, $\sim$150 fs) is reduced in repetition rate to 40~MHz by an acoustic-optic modulator (AOM). The electro-optic modulator (EOM0) and PBS are used to adjust the pump energy of each pulse. This controls the squeezing degree ($r_{i}$) of the squeezed vacuum states in each time-bin. The spectral mode of the pump light is modulated by a spatial light modulator (SLM), two gratings and cylindrical lens (CL). Then, spectrally uncorrelated two-mode squeezed light can be generated by pumping the periodically poled KTP (ppKTP) waveguide~\cite{Eckstein2011,Harder2016,Bell2019}. Following interference at a 50:50 beamsplitter (BS), a series of individually addressable single-mode squeezed states (SMSSs) can be efficiently prepared~\cite{Lvovsky2016}.
(2) Quantum processing unit. The SMSSs are then sent into a time-bin interferometer, which is programmed for a specific unitary operation. This is achieved according to Clements' architecture~\cite{Clements2016}, which is realized by a group of Mach-Zehnder interferometers (MZIs) consists of two fast optical switches EOMa and EOMb, a 7.5 m delay line (to combine or separate two adjacent time bins), and a linear transformation $T(\theta,\varphi)$ achieved by EOM1 and EOM2. Since the optical path before and after $T(\theta,\varphi)$ pass through the same low-loss free-space delay line, the phase stability of the setup is well guaranteed, and the non-uniform loss expected in the fiber-loop scheme~\cite{Motes2015} is mitigated.
(3) Quantum sequential access memory (QuSAM). In each loop of evolution, the quantum memory is achieved by a 180-meter-long optical fibre delay line. The QuSAM ensures that the last time-bin has completed the operation in one cycle before the first time bin enters into next cycle. With a 4f beam-shaper system, we can efficiently couple the light from free space into single-mode optical fiber, and realize a low-loss time-bin memory (with total efficiency of $\approx94\%$) by reshaping the spatial mode of the beam.
(4) Detection module. The experiment is required to be run with collision-free detections. After the linear transformation, the photonic time-bin modes are sent into superconducting nanowire single photon detector (SNSPDs), and the output photons on each time-bin can be measured.

As illustrated in Fig. 1(a), this time-bin encoded GBS machine enables us to expand the number of modes arbitrarily, and freely set the required squeezing parameters and linear transformation matrix for the tasks with a series of EOMs. Thus, this universal and programmable architecture supports arbitrary GBS circuits to be run on this machine. As a concrete example, benefiting from these merits, the adjacency matrix $A$ of a graph $\mathcal{G}$ can be encoded into this GBS machine by decomposing $\mathcal{L}(A)$ (Laplacian of graph $\mathcal{G}$) after a suitable rescaling, as shown in the inset of Fig. 1(a).

The validation of \emph{Abacus} can be demonstrated by the sampling results from two random GBS circuits with different dimensions. The normalized photon sampling distribution probabilities are shown in Fig. 1(b) and (c). In Fig. 1(b), a 32-mode random interferometer is chosen, and only four squeezers are turned on ($r_{1-3,32}=2.23$) here. The statistical results of all two-photon detection events are plotted, and the total variation distance (TVD) between experimental and theoretical results is 0.054. Similarly, the four-photon distribution pattern is shown in Fig. 1(c), which is carried out on a 16-mode GBS with all 16 squeezers turned on and $r_{\text{max}}=1.8$ (here, TVD is 0.175). We also use the modified likelihood ratio test introduced in~\cite{Spagnolo2015} to exclude the thermal state and distinguishable photon hypotheses, and these details can be found in Supplementary Information II.G. These show that \textit{Abacus} can perform the sampling tasks with high fidelity.

\emph{Finding the maximum weighted clique with GBS}---Not only can GBS be used to demonstrate quantum advantage in the laboratory~\cite{Zhong2020,Madsen2022}, as a near-term specific-function quantum computer, it can also be used in solving certain problems in real-world applications. Here, we first use \textit{Abacus} to solve the max clique decision problems, which are NP-hard problems in graph theory, and plays a crucial role in many applications~\cite{Karp1972}.

Clique refers to all the maximal complete subgraphs in a graph $\mathcal{G}$, and clique-finding is a problem with a complexity which scales exponentially with the number of nodes. Here, we use \emph{Abacus} to find the maximum weighted clique in a graph. A 32-node weighted graph $\mathcal{G}_{32}$ is artificially constructed here (details are shown in Supplementary Information IV), and the essential step is encoding $\mathcal{G}_{32}$ onto our GBS machine. Using the method introduced in Ref.~\cite{Bradler2018,Banchi2020}, we perform Takagi-Autonne decomposition to the Laplacian of graph $\mathcal{G}$ with appropriate re-scaling, and obtain the unitary operation $U$ and squeezing parameters $r_{i}$ which are required to be programmed on the GBS (see the Methods for details). Then, we control the AOM chopper with an arbitrary waveform generator (AWG) to pump the ppKTP waveguide with 32 sequential pulses. EOM0 is used to adjust $r_{i}$ for each time-bin, and $U$, the unitary operation, is achieved by adjusting the input voltages of EOM1 and EOM2 in the time-bin interferometer. After the mapping $\mathcal{G}_{32}$ onto \textit{Abacus}, around 300 five-(or more)-photon sampling results are then collected in each experiment. Using these sampling data, we can find the cliques with nodes corresponding to the 30-time post-processed sampling results (see details in Methods). Fig. 1(d) displays all six-node cliques and their corresponding probabilities. The maximum weighted clique stands out as the most probable among them. In comparison to classical sampling with the same post-processing iterations, GBS demonstrates a significantly higher probability of successfully finding the maximum weighted clique, approximately twice as much. This indicates that GBS can perform the clique-finding task with high efficiency~\cite{Banchi2020}.

\begin{figure*}[ht]
\centering
\includegraphics[width=1.00\textwidth]{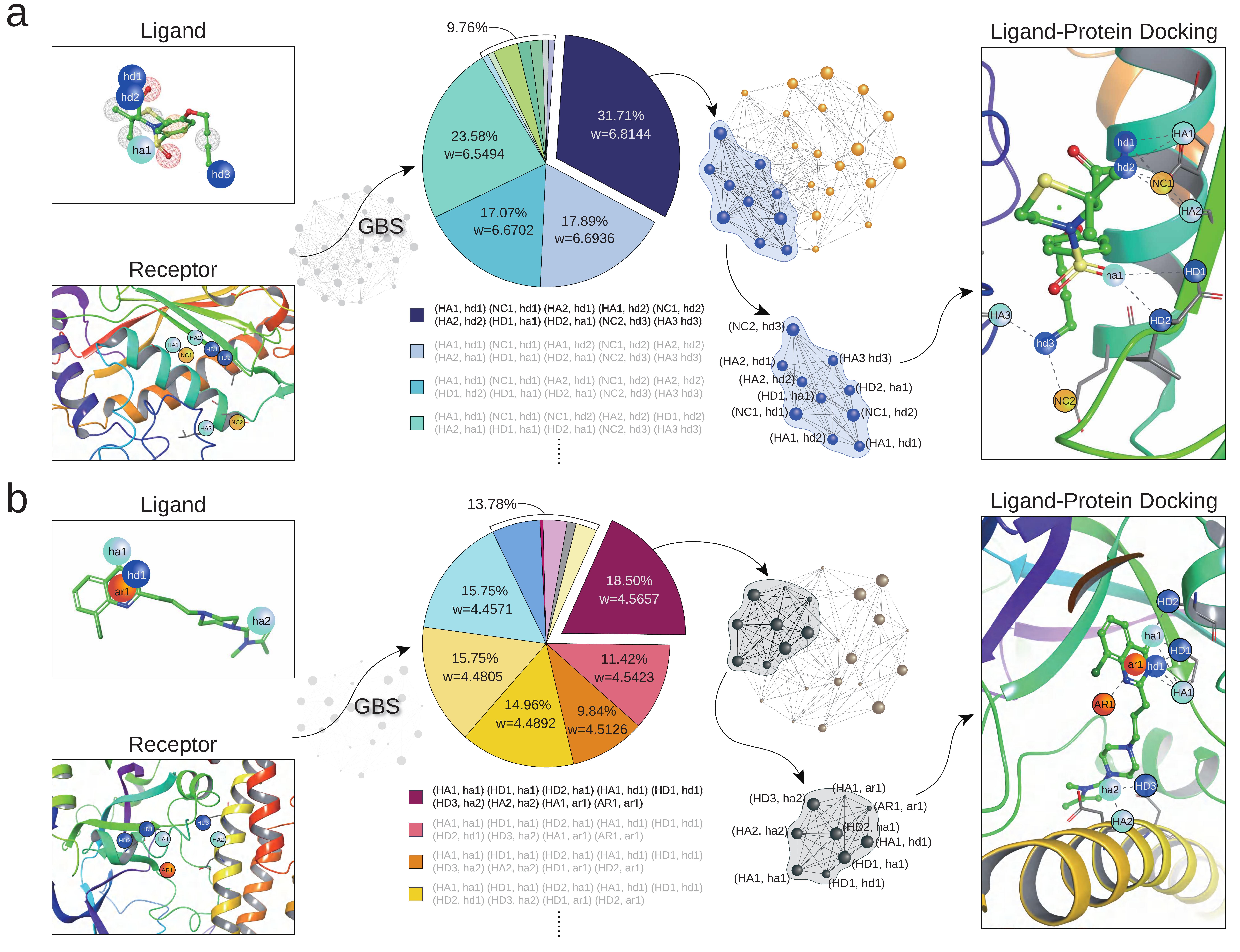}
\caption{\textbf{Molecular docking results obtained with \textit{Abacus}}. \textbf{a,} The docking pair of PARP-CQ. \emph{Abacus} is encoded with a 24-node BIG for finding the maximum weighted clique. 347 sample data and 100-iteration local searches are used. \textbf{b,} Another 28-node BIG constructed by the complex of TACE-TS. 254 sample data and 10 iterations are applied here.
The colored spheres denote the pharmacophore points we considered in the experiment (red: hydrogen-bond acceptor (HA), blue: hydrogen-bond donor (HD), yellow: negative charge (NC), orange: aromatic (AR), and we use capital (lowercase) letters to represent pharmacophore points in the protein (ligand)). The sphere meshes in the ligands are the other possible pharmacophore points but are not considered in our experiments. All the cliques found from GBS experiments are shown in the middle pie charts, and the maximum weighted cliques in both cases are shown with a major proportion in the experimental results obtained from QIVS. More details are found in the Supplementary Information V, VI.}
\label{Fig2}
\end{figure*}

\begin{figure*}[hbt]
\centering
\includegraphics[width=1.00\textwidth]{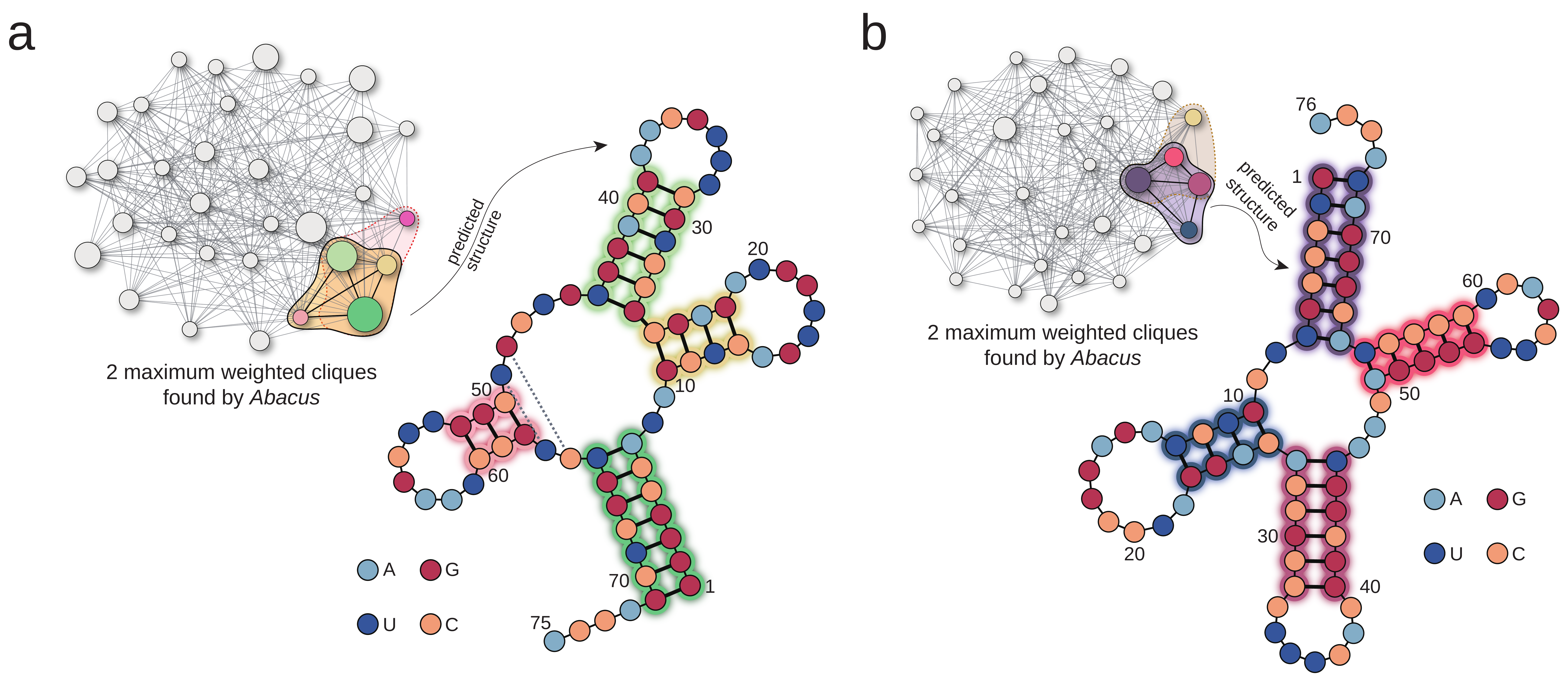}
\caption{\textbf{RNA folding prediction results}. The GBS-based RNA folding prediction results of two RNA sequences (\textbf{a}: Accession Number: AH003339, and  \textbf{b}: Accession Number: AB041850). The grey dashed lines represent the false negative base-pair matching. The four colored circles (A, G, U, C) in the RNA structure represent four different bases. The color of nodes in cliques corresponds to the predicted stems in RNA folding structure (i.e., the color of shadows at the corresponding base-pairs). More details are shown in the Supplementary Information VII.}
\label{Fig3}
\end{figure*}

\emph{Molecular docking with GBS}---If the graph is constructed according to an actual system occupying the network structure, the clique-finding task then could be utilized to find the optimal subset corresponding to the maximum weighted clique. Recent research shows that the information of the best docking orientation of the protein-ligand complex can be predicted by the maximum weighted clique of binding interaction graph (BIG), which is a weighted graph constructed based on docking modes between ligand and receptor~\cite{Banchi2020}. In the BIG, the weighted nodes represent the interacting pharmacophore pairs weighted by potential, and the edges represent the compatible contacts (see the Methods). By encoding the BIG on \emph{Abacus}, we can solve molecular docking problems by finding the maximum weighted clique in BIG~\cite{Kuhl1984,Banchi2020} as we demonstrated in Fig. 1(d).

In order to better demonstrate the capability of GBS in solving molecular docking problems, we build a quantum inverse virtual screening (QIVS) platform based on \emph{Abacus}, and use two pairs of protein-ligand complexes with different drug properties to demonstrate the ability of QIVS in drug design and verified the practicability of the platform. A 28-node BIG $\mathcal{G}_{\text{PARP-CQ}}^{28}$ is constructed based on the Poly (ADP-Ribose) polymerase-1 (PARP) and an 8-chloroquinazolinone-based inhibitor (PARP-CQ), which is a promising candidate for anti-cancer drugs~\cite{Powell2010,Lu2018,Tangutoori2015} or some central nervous system (CNS) diseases such as Parkinson's and Alzheimer's diseases~\cite{Olsen2019,Martire2015}. The structures of ligand and protein and their BIG are shown in Fig. 2(a). By encoding $\mathcal{G}_{\text{PARP-CQ}}^{28}$ onto \textit{Abacus}, we collect the sampling results and find the associated cliques with post-processing (i.e., shrink and local search)~\cite{Banchi2020}. The pie-chart in Fig. 2(a) shows all the cliques we find with GBS experiments, where each sector corresponds to the different cliques with corresponding weight. It is clear that the maximum weighted clique (with seven nodes and weight~=~6.8144) occupies the major proportion and the proportions increases with the clique weights. It demonstrates we can use \textit{Abacus} to find the best binding pose (Fig. 2(a), right side) of this complex with a high success rate.

In the second case, we use the complex of tumour necrosis factor (TNF)-$\alpha$ converting enzyme (TACE) and thiomorpholine sulfonamide hydroxamate inhibitor (TACE-TS)~\cite{Levin2006}, which are involved in inflammatory diseases~\cite{Chemaly2017}. Aromatic pharmacophore points are included here, and in order to increase the accuracy of the docking results, an improved algorithm is used here. Considering the fact that the interaction strength between different pharmacophore points may exist some difference, in this case, the variable distance is used to compare the distance between different points when we construct this 24-node BIG $\mathcal{G}_{\text{TACE-TS}}^{24}$. GBS experiments then is performed by programming the circuit with another set of $\theta_{i}$, $\varphi_{i}$ and $r_{i}$. There are a total of 11 cliques that we find in the sampling results, and around six of them (maximum size $N=9$) appear with relatively high probabilities. The ligand-protein docking position suggested by the maximum weighted clique (with weight~=~4.5657) is shown on the right in Fig. 2(b). Compared with the method in Ref~\cite{Banchi2020}, the improved method we proposed here does get a more accurate binding pose result, and the detailed comparison analysis is shown in the Supplementary Information V.

Although there is relatively high loss in the experiment, the maximum weighted clique can still be found with a high success probability through post-processing, which is robust to noise~\cite{Banchi2020}. The above results, predicted by GBS experiments, agree well with the outcomes obtained from the corresponding co-crystal structure, which can be found by reviewing complex structures (PDB ID 2A8H, 7ONR) within a certain distance ($\tau$) to each other~\cite{Levin2006,Johannes2021,Berman2003}.

\emph{GBS for RNA folding prediction}---The molecular docking process heavily relies on the protein structure, and the fact is that many pathogenic proteins associated with human diseases cannot be targeted by conventional small-molecule drugs or biomacromolecules~\cite{Santos2016}. In recent years, nucleic acid drugs have gained attention in the pharmaceutical field as a potential solution to overcome the limitations of existing target drugs and to treat previously ``untargete'' diseases. Predicting RNA structures has become an important task in discovering these nucleic acid drugs, as it can aid in identifying potential drug targets and predicting small molecule drugs' interactions with RNA molecules~\cite{Townshend2021}. However, predicting RNA structures by calculation has proven difficult, as only a few RNA structures are known. Nevertheless, exciting works in protein and RNA structure prediction have emerged recently, with artificial intelligence technology being particularly prominent~\cite{Townshend2021,Jumper2021}. Quantum computational technology also has great potential to solve this folding prediction task~\cite{Emani2021,Zaborniak2022}. However, this problem has not yet been experimentally demonstrated on devices that can exhibit quantum computational advantages (such as GBS devices) due to their programmability limitations.

Using our universal programmable GBS device \emph{Abacus}, we use a new method, inspired by Ref.~\cite{Tang2022}, for predicting RNA sequence folding. This approach involves modeling the RNA sequence as a weighted full stem graph (WFSG) and then encoding it into our universal programmable GBS device. The WFSG captures all possible folding information of the RNA sequence, where each node represents a possible stem in the sequence, and the edges indicate the co-existence between them~\cite{Tang2022}. The weight of each node corresponds to the length of the stem it represents. Then, the RNA folding prediction can be obtained by finding the maximum weighted cliques in WFSG~\cite{Tang2022}. To demonstrate the effectiveness of our GBS machine in solving this problem, we conducted two experiments with different RNA fragments on \emph{Abacus}, and the results are shown in Fig. 3.

In the first example, we predicted the secondary structure of an RNA sequence (Accession Number: AH003339) by encoding the corresponding 32-node WFSG into \emph{Abacus}. We found a total of two maximum weighted cliques, and the Matthews correlation coefficient (MCC) of the best one (shown in yellow shadow in Fig. 3(a)) reached 0.953, which outperforms FOLD (best case) and RNAProbing, with MCC values of only 0.864 and 0.934, respectively. In the second experiment, we use the RNA sequence of the organism Alanine (Accession Number: AB041850) and encoded its corresponding WFSG, which had 31 nodes, into \emph{Abacus} by modifying the control program. The best prediction with MCC=1.00 among the two results is shown in Fig. 3(b), and it is more accurate than those obtained by other methods, with MCC values of $\text{MCC}_{\text{FOLD}}=0.870$ and $\text{MCC}_{\text{RNAProbing}}=0.914$. Details of the true reference folding and other information are provided in the Supplemental Information VII.

\emph{Discussion and outlook}---The scalability and programmability of our universal GBS machine enable its utilization in real-world applications, as demonstrated in this work. The ability to program arbitrary graphs demonstrates that drug discovery tasks, such as molecular docking or RNA folding prediction, can be performed efficiently by a purpose-built quantum computer. However, unequivocal quantum computational advantage~\cite{Zhong2020,Madsen2022} has not been realized in our experiments due to photon loss. Although the question of whether GBS can outperform improved classical algorithms or quantum-inspired algorithms remains open~\cite{Oh2022}, and the potential for GBS to demonstrate computational advantages also relies on the properties of the encoded graph, we remain optimistic about scaling \textit{Abacus} to several hundred modes using the ``multi-core encoding'' and ``distributed computing'' methods. This scalability holds the potential to unlock quantum advantages in some specific real-world applications. Additionally,  it is crucial to consider practical applications that encompass more complex protein structures, larger pharmacophore points, and longer RNA sequences, which also necessitate the use of such a large-scale GBS machine. For a comprehensive discussion on scaling our GBS machine by minimizing loss and utilizing the ``multi-core encoding'' and ``distributed'' encoding method, please refer to Supplemental Information VIII.

With the help of time-bin encoding and fast computer-controlled EOMs, we have successfully constructed \emph{Abacus}, a universal and programmable GBS machine which allows for arbitrary setting of the scale and squeezing level of each mode. The interferometer also supports arbitrary unitary operations, making it a state-of-the-art photonic quantum processor. By utilizing this device, we have demonstrated the advantage of GBS over classical sampling in the clique finding task for a 32-node weighted graph. Moreover, we have developed a QIVS platform by utilizing GBS in place of the classical docking process, and we have applied it to perform GBS molecular docking for two different types of molecules with either TACE or PARP1 proteins. The sampling results effectively indicate the maximum weighted clique and demonstrate the ability of our GBS machine to search for the optimal docking structure. Furthermore, successful RNA folding prediction tasks have been accomplished on \emph{Abacus}, which contributes to the advancement of nucleic acid drug development. Apart from offering programmability and universality, this work presents a promising hardware solution for the near-term industrial implementation of quantum computing in the biopharmaceutical industry. It also paves the way for diverse real-world applications in the future.

\textbf{Data availability}\\
Source data for Figs. 1(b), (c), and (d) are available at xxx. The graph information used in Figs. 2 and 3 is available with this paper.

\textbf{Code availability}\\
The codes used to generate the corresponding matrices in Fig. 2 and Fig. 3, analyze experimental data, and implement the Bron-Kerbosch algorithm and Maximum Clique algorithm for result verification can be obtained from xxx.

\textbf{Methods}

\textbf{Mapping a graph onto GBS}\\
For the node-weighted graph $\mathcal{G}$, we can define a matrix $B=\Omega(D-A)\Omega$, where $A$ is the adjacency matrix of $\mathcal{G}$, $\Omega$ is a diagonal matrix with elements $\Omega_{ii}=c(1+\alpha \omega_{i})$  and weights $\omega_{i}$, and $D$ is the degree matrix of $A$ defined as $D_{ii}=\sum_{j}A_{ij}$. In order to find suitable squeezing parameters of the input states, and guarantee the spectrum of $B$ within $[0,1)$, we need to carefully choose the parameters $c$ and $\alpha$ in experiment. Based on the properties of $A$, it is clear that the matrix $B$ is a symmetric matrix, and we can decompose it into $U\oplus^{N}_{i}\tanh(r_{i})U^{T}$. Here, $r_{i}$ are the squeezing parameters we need to adjust in the experiments, and $U$ can be programmed with the time-bin encoded interferometer. Then, we can detect the cliques from the sampling results with the help of post-processing~\cite{Banchi2020}. A more detailed explanation is presented in Supplementary Information.

\textbf{Constructing the adjacency matrix of BIG}\\
Assuming $n$ and $m$ pharmacophore points are considered in the ligand and the potential protein, respectively, then the adjacency matrix of BIG ($\mathcal{G}_\text{BIG}$, Fig. S5.1) will be a $nm$$\times$$nm$ symmetric matrix $\mathcal{A}$, with diagonal elements that are all zero. Other elements are determined by the distances of the pharmacophore pairs in the ligand and the protein (the dashed lines in boxes 3 and 4, as shown in Fig. S5.1), respectively. As for the $\mathcal{A}_{ij}$, if the distance difference between corresponding pharmacophore pair in ligand ($D_{\text{L}}$) and protein ($D_{\text{P}}$) does not exceed $\tau+2\varepsilon$ (where $\tau$ is the flexibility constant, and $\varepsilon$ is the interaction distance), i.e., $D_{\text{P}}-D_{\text{L}} < \tau + 2\varepsilon$, we set $\mathcal{A}_{ij}$ and $\mathcal{A}_{ji}$ to 0. Otherwise, we set them as 1.

For more accurate docking results, when the interaction between each pharmacophore points is different, we trial different interaction distances $\varepsilon$ to determine the elements in BIG. For example, we use $\varepsilon_{1}$ to determine the group when all pharmacophore points are hydrogen-bond acceptor and donor and use $\varepsilon_{2}$ to determine the group when the pharmacophore points contain hydrogen-bond acceptor/donor and aromatic (or other pharmacophore points which will induce different interaction strengths). More details and distance data are shown in the Supplementary Information.

\textbf{The programmable GBS machine}\\
Here we provide more details about the setup in Fig. 1(a):
(1) Squeezed state preparation: squeezed light is generated by pumping a 10~mm long periodically-poled potassium titanyl phosphate (ppKTP) waveguide. The pump light at 773 nm is produced from a Ti:sapphire laser, and is spectrally tailored for producing 1.9 nm pulses. Two-mode squeezed states (TMSS) can be converted to single mode squeezed state (SMSS) with the same degree of squeezing using a 50:50 BS. (2) Quantum processing unit: An MZI is realized with EOMa/b and a 7.5~m long free-space delay line, and EOM1/2 are applied to adjust $\theta$ and $\varphi$. Any arbitrary unitary operation can be executed based on Clements' architecture~\cite{Clements2016}. (3) QuSAM: a 180~m long single-mode fibre serves as an optical delay line to store the time-bins and ensures that the sequence can evolve in a predetermined order. (4) Detection: SNSPDs are used to detect the single photon events since our experiments are performed in the collision-free space. To avoid the issue of the SNSPD dead-time ($\lesssim 50$~ns) being longer than the time interval between two adjacent time bins (25~ns), we use another EOM to separate two adjacent time bins and use two SNSPDs for detection. The throughput of each round-trip in the system is approximately 82\%. In the case of Fig. 1(b), the average count rate of two-fold events is 45 counts per second, and in the case of Fig. 1(c), the four-fold average count rate is 24 counts per second, which are calculated from $10^{7}$ samples in 10 minutes (the repetition rate of each individual sampling experiment is 20 kHz). More details are provided in the Supplementary information.

This GBS machine has two main advantages compared with previous works:

\paragraph{$\bullet$} Universal operation is possible where both the squeezers and arbitrary unitary matrices can be programmed on the time-bin interferometer. This makes is suitable not only for molecular docking of various molecules but also for other applications. Our architecture also provides flexibility in scaling to many modes via the control software. Compared to previous work~\cite{Zhong2020,Madsen2022}, our GBS setup supports adjustments to all the parameters: $n$ squeezing parameters $r_{i}$ and $n(n-1)/2$ parameters for an arbitrary $U$. Our time-bin-encoding GBS setup is resource efficient for scaling up. Specifically, when the number of modes increases, we do not need to add more squeezed light sources. Independent of the number of modes we required in experiments, two analog EOMs (i.e., EOM1 and 2) assisted with two light-switch EOMs (i.e., EOMa and b) are sufficient to realize any linear transformation. A resource advantage is also exhibited in the detection. As we discussed in the main text and Supplementary Information section II. E, two SNSPDs are enough for collecting the $\sim30$-mode GBS samples.
\paragraph{$\bullet$} Non-uniform loss in previous time-bin interferometer implementations appears across different time-bin modes, and this limits the ability to perform an arbitrary unitary operation~\cite{Motes2015}. In this setup, we use a free-space delay-line with transmittance 0.995 to greatly reduce the non-uniform loss. Thus, the mitigated non-uniform loss and dispersion-free features in our setup can better exhibit universality. The time-bin encoded GBS scheme is intrinsically phase stable~\cite{He2017}. As shown in Fig. 1, since every time bin goes through the same path, the slow phase fluctuations (caused by mechanical vibrations, temperature drifts, or other unpredictable environmental noise) can be neglected compared to the fast sampling rate where a sample is obtained in 50 $\mu$s. The 7.5~m free-space delay-line is isolated from the environment to ensure that the phase between two adjacent time-bins can be stable for up to five minutes. This is enough for collecting $10^{6}$ samples within one minute.

\textbf{Post-processing method with ``Shrinking'' and ``Local Search''}\\
The presence of various types of noise in the experiments affect the probability of the maximum weighted clique obtained from the raw experimental data. In some cases, the subgraph obtained may not even be a clique. Certain types of noise are unavoidable in experiments (e.g. photon loss) in which case we can use the raw experimental data as a seed which can be input to a post-processing algorithm to generate cliques at a high-rate. The post-processing method introduced in Ref.~\cite{Banchi2020} is very useful for this purpose. We briefly review the post-processing method here and discuss how we use in our experiments.

First, we use ``Greedy Shrinking'' to ensure the subgraph obtained from the raw GBS data is a clique by removing the nodes based on the degree and weight of the nodes until it forms a clique. In order to obtain the maximum weighted clique, which usually occurs with a larger number of nodes, we perform an expansion with a ``Local search''. This expands the clique by adding neighboring nodes within several iterations of the algorithm to generate the largest clique. This is represented as the ``Post-processing'' module in Fig. S5.1. Strawberry Fields ~\cite{Killoran2019} is used to perform the post-processing. Further details can be found in Ref.~\cite{Banchi2020}.

\textbf{Further scaling by reducing loss}\\
In our experiment, simultaneously achieving universality and programmability comes at the cost of loss, which increases with the circuit depth or more specifically the number of cycles. This relatively large loss exists in our GBS machine, and prohibits demonstration of quantum computational advantage. Although we mainly focus on the mapping of GBS to real-world applications, with further developments towards low-loss optical components quantum advantage should be possible in the future.

Loss in \emph{Abacus} can be reduced by various methods. Loss in the experiment mainly comes from: 1) the coupling loss from ppKTP waveguide to single-mode fibre; 2) insertion loss caused by EOMs; 3) the limited coupling efficiency from free space to QuSAM; and 4) the limited detection efficiency of SNSPDs.
Particularly, for 2) and 3), due to the characteristics of our free-space loop architecture, the total loss will increases exponentially with the loss inside the loop. Therefore, when the number of modes is large, small improvements to these sources of loss will greatly improve the overall loss.

Firstly, for the coupling loss from the ppKTP waveguide to single-mode fibre, mode shaping techniques may be applied to match the spatial mode of light from the ppKTP waveguide to that of the fibre, potentially improving the coupling efficiency to greater than $0.9$~\cite{Sun2021}. Secondly, the insertion loss of EOMs or other optical elements is unavoidable. However, an EOM with a shorter, lower loss, crystal (driven by a higher gain amplifier) could be used. Combining the actions of EOM1 and EOM2 into EOM operation will further reduce loss experienced upon reflection and absorption at the end faces. Transmission has been shown to reach higher than $0.99$ after optimizing the EOM~\cite{Madsen2022}. Thirdly, the coupling efficiency from free-space to QuSAM can be improved up to $\approx0.97$ by using 4f- or 8f-imaging systems through spherical lenses and graded-index lens fibre couplers (see Ref.~\cite{Madsen2022}). In addition, a Herriott long-distance delay line can be used as a quantum memory for minimizing the loss~\cite{Enomoto2021}. Finally, the detection efficiency can be improved with the latest generation of nanowire detectors with detection efficiencies at 1550 nm of up to $0.95$. Through these methods to minimize the loss, the single-loop transmission can potentially reach $\approx90\%$. Thus, our GBS setup can be extended to at least 60 modes (e.g., a 60-mode GBS experiment will have total transmission $0.90^{61}$$\times0.9$ (coupling efficiency from ppKTP to fibre)$\times$0.944 (filter after ppKTP induce)$\times$0.93 (coupling efficiency from QPU to fibre)$\times$0.973 (transmission of demultiplexer)$\times$0.95 (detection efficiency of SNSPDs)$\approx$$0.12\%$, which is greatly improved compared with the experiments shown in Fig. 1(a)).

\textbf{Implement displacement and photon-number resolving detection in time-domain GBS}\\
Displacement operation $D(\alpha)$ is entirely feasible to include in \textit{Abacus}. To achieve this, the photon source module needs to be rebuilt, which involves incorporating an optical parametric oscillator (OPO) after the mode-locked laser (Chameleon) to generate a 1550nm laser. This laser is subsequently split into two separate paths. In one path, the light serves as the pump for generating squeezed states, similar to our previous work. In the other path, the light is utilized as a coherent state to achieve the displacement operation. The addition of a delay line ensures that the coherent state and squeezed state reach the BS simultaneously, enabling optimal interference at the output. Two EOMs facilitate the programmability of the amplitude and phase of the displacement operation. For more detailed information, please refer to Supplemental Information Sec. II.H.

Our time-domain GBS machine \textit{Abacus} can also implement photon-number resolving detection with transition edge sensor (TES). The TES initially needs to be cooled below its transition temperature of approximately 100 mK and then heated back to its transition temperature by applying a bias current~\cite{Irwin1995}. To maintain the TES at this temperature, it should be operated inside a dilution refrigerator. After a photon absorption event, it takes approximately 5 $\mu$s for the TES to return to its original temperature. Therefore, the repetition rate of TES detectors is usually limited to around 100-300 kHz. This necessitates the installation of a demultiplexer in our time-domain GBS setup. To address this limitation for time-domain GBS machine, a demultiplexer needs to be installed here. By employing a loop structure, it is straightforward to implement a 9-channel demultiplexer using a single electro-optic modulator (EOM). The EOM enables us to manipulate the polarization of photons in each time-bin, thereby determining whether they exit the system through polarization beam splitters (PBS) or undergo internal reflection within the loop. The design corresponding to this approach is depicted in Fig. S2.14, and further detailed information can be found in Supplemental Information Sec. II.I.

\textbf{Pharmacophore points selection }\\
In general, the selection of PARP-PARPi pharmacophore points is according to previous research focused on PARP-PARPi relationships~\cite{Kumar2019,Zandarashvili2020}. These articles have demonstrated important amino acid residues from the protein and functional groups from the inhibitor that will influence the efficacy of the protein-ligand interactions. We choose some of them for the GBS machine due to the size of the experiment.

The selection of pharmacophore points in this work is typically based on prior knowledge from experimental studies, structural analysis, or computational modelling of similar PARP-PARPi complexes. The selection of pharmacophore points, such as Hydrogen-bond acceptors/donors, Negative charges, Pi-Pi interaction, and aromatic ring, in the PARP-PARPi complex is based on their known importance in the interaction between the protein (PARP) and the ligand. These pharmacophore characteristics play a crucial role in the binding affinity and specificity of ligands to protein targets.

Hydrogen-bond interactions are important for stabilizing the ligand-protein complex. Such an interaction occurs between the hydrogen atom of the ligand and a hydrogen bond acceptor or donor group on the protein. These interactions contribute to the overall binding strength and specificity by forming specific and directional interactions. Pi-Pi interactions involve the stacking of aromatic rings in the ligand and the protein. These interactions are driven by the pi electrons present in the aromatic systems and contribute to the stability of the complex. Pi-Pi interactions are often found in ligand-protein interactions and can enhance binding affinity. Some articles have reviewed pharmacophores in a PARP inhibitor and the nicotinamide component is considered as a Hydrogen-bond donor and acceptor, as well as a part of Pi-Pi interaction with the tyrosine residue~\cite{Kumar2019}.

Aromatic rings are frequently present in ligands and proteins and can participate in various types of interactions, including Pi-Pi stacking, hydrophobic interactions, and van der Waals interactions. Aromatic rings provide a hydrophobic surface that can interact with complementary hydrophobic regions in the protein, contributing to the overall binding affinity. The aromatic ring at the tail of the compound is also critical, for which we take it as a pharmacophore as well~\cite{Johannes2021}.

Negative charges, represented by negatively charged atoms or functional groups, also play a significant role in the PARP-PARPi complex. Such a negative charge can interact with positively charged residues on the protein, like arginine or lysine, through electrostatic interactions, and these interactions can alsocontribute to the stability of the complex and enhance ligand binding.

\textbf{Acknowledgements} S. Yu would like to express special gratitude to Prof. Sung Ha Kang and Dr. Mengyi Tang for their assistance and support in RNA sequence folding prediction. We thank Qin-Qin Wang, Bryn Bell, Zhian Jia, Xiao-Ye Xu, Zheng-Hao Liu and Aonan Zhang for the helpful discussions, and Santiago Sempere-Llagostera for providing useful feedback on the manuscript.
This work was supported by Engineering and Physical Sciences Research Council and Quantum Computing and Simulation Hub (T001062), UK Research and Innovation Future Leaders Fellowship (MR/W011794/1), EU Horizon 2020 Marie Sklodowska-Curie Innovation Training Network (No. 956071, ``AppQInfo''), Innovation Program for Quantum Science and Technology (No. 2021ZD0301200), International Postdoctoral Exchange Fellowship Program 2022 (No. PC2022049), China Postdoctoral Science Foundation funded projects (No. 2020M681949), the National Natural Science Foundation of China (Nos. 11821404, 12174370 and 12174376), the Youth Innovation Promotion Association of Chinese Academy of Sciences (No. 2017492), the Fok Ying-Tong Education Foundation (No. 171007).

\textbf{Author contributions} S.Y., and J.S.T. designed the experiment with input from R.B.P. S.Y. built the experimental setup with the help of Q.P.L., R.B.P., L.X., Y.M., Z.P.L., Y.Z.Y., Z.A.W., N.J.G., W.H.Z., and Y.T.W. Z.P.Z. developed the control system for the programmable hardware. Y.F. and G.K.T. designed the molecular docking scheme. S.Y., and Y.T.W. designed and built the squeezed-light source with the help of R.B.P. S.Y., and Y.F. completed data analysis with the help of W.L., J.S.T., R.B.P., and I.A.W. J.S.T., Y.T.W., S.Y., S.S.-S., and Y.D. designed the further scaling method with the help of Z.L. S.Y. and Y.F. prepared the manuscript with the help of R.B.P., J.S.T., Y.T.W., Z.L., E.M., S.E.T., Y.D., C.F.L., I.A.W., and G.C.G. All authors discussed the results and reviewed the manuscript.

\end{document}